\documentclass[notitlepage,pra]{revtex4-1}

\usepackage{inconsolata}
\usepackage[strict,uselistings,pretty]{revquantum}
    \definecolor{dgreen}{rgb}{0,.4,0}
    \definecolor{dblue}{rgb}{0,0,.7}
    \newaffil{MSRQuArC}{\smallskip
        Quantum Architectures and Computation Group,
        Microsoft Research,
        Redmond, WA 98052, United States
    }
    \lstdefinelanguage{QSharp}{%
        language     = Java,
        morekeywords = {namespace,operation,function,body,let,mutable,repeat,until,fixup,newtype,fail,using,borrowing,set,elif,adjoint,controlled,auto}
    }
\usepackage{letltxmacro}
    \newcommand*{\SavedLstInline}{}
    \LetLtxMacro\SavedLstInline\lstinline
    \DeclareRobustCommand*{\lstinline}{%
    \ifmmode
        \let\SavedBGroup\bgroup
        \def\bgroup{%
        \let\bgroup\SavedBGroup
        \hbox\bgroup
        }%
    \fi
    \SavedLstInline
    }

\newcommand{\qs}{Q\#}
\newcommand{\LIQUID}{{LIQ{\em Ui}$|\rangle$}}

\begin{document}

\title{\qs: Enabling scalable quantum computing and development\\ with a high-level domain-specific language}

\author{Krysta~M.~Svore}
    \affilMSRQuArC
%    \email{ksvore@microsoft.com}
\author{Alan Geller}
    \affilMSRQuArC
%    \email{alan.geller@microsoft.com}
\author{Matthias Troyer}
    \affilMSRQuArC
%    \email{mtroyer@microsoft.com}
\author{John Azariah}
    \affilMSRQuArC
%    \email{johnaz@microsoft.com}
\author{Christopher Granade}
    \affilMSRQuArC
%    \email{christopher.granade@microsoft.com}
\author{Bettina Heim}
    \affilMSRQuArC
%    \email{bettina.heim@microsoft.com}
\author{Vadym Kliuchnikov}
    \affilMSRQuArC
%    \email{vadym@microsoft.com}
\author{Mariia Mykhailova}
    \affilMSRQuArC
%    \email{mamykhai@microsoft.com}
\author{Andres Paz}
    \affilMSRQuArC
%    \email{anpaz@microsoft.com}
\author{Martin Roetteler}
    \affilMSRQuArC
    \email{martinro@microsoft.com}

\begin{abstract}
    Quantum computing exploits quantum phenomena such as superposition and entanglement to realize a form of parallelism that is not available to traditional computing.
    It offers the potential of significant computational speed-ups in quantum chemistry, materials science, cryptography, and machine learning.
    The dominant approach to programming quantum computers is to provide an existing high-level language with libraries that allow for the expression of quantum programs.
    This approach can permit computations that are meaningless in a quantum context; prohibits succinct expression of interaction between classical and quantum logic; and does not provide important constructs that are required for quantum programming.
    We present \qs, a quantum-focused domain-specific language explicitly designed to correctly, clearly and completely express quantum algorithms.
    \qs~provides a type system, a tightly constrained environment to safely interleave classical and quantum computations, specialized syntax, symbolic code manipulation to automatically generate correct transformations of quantum operations, and powerful functional constructs which aid composition.
\end{abstract}

\maketitle

%=============================================================================
\section{Introduction}
\label{sec:intro}
%=============================================================================

While classical computing models a binary system and employs boolean logic for computation, quantum information is stored in quantum states of matter, and computation is effected by operations based on quantum interference.
Just as individual binary values are stored in bits, quantum states are stored in qubits.
A quantum operation operates on one or more qubits and is called a gate.
A sequence of gates operating on a set of qubits is traditionally called a circuit.
Most quantum programming languages are effectively circuit descriptions.
Quantum algorithms may employ patterns and constructs that are not readily expressible in circuits.
Constructs such as classical control predicated on the result of a quantum measurement are difficult to express as circuits, while recursion and unbounded iteration are often impossible.
Further, the physics and engineering of quantum computing devices currently requires that many low-level concerns such as error-correction, noise and timings need to be explicitly addressed.
However, a software developer requires these details to be abstracted away so that focus is maintained on expressing a higher-level quantum algorithm.

To address the need for expressivity and abstraction, we developed a special language which provides a clean separation between the code to be executed in the quantum context from the driver code that can be programmed in a traditional high-level language.
Additionally, since only some hard computational parts of a ``program'' are quantum mechanical in nature, it is helpful to consider a quantum computer not as a full-fledged computation device in its own right (replete with mechanisms for storage, network communication, user interaction and so on), but rather as an external, adjunct, co-processor to a classical machine.
In such a model, the quantum computation happens on the co-processor in a cleanly separated domain, and indeed, becomes the domain for which we design a domain-specific language (DSL) such as \qs.
Finally, since many quantum operations are, in fact, unitary transformations, it is quite a common problem in quantum computing to require the adjoint (formally, the conjugate transpose) of a given transformation.

Without symbolic computing, the only way to produce the adjoint of a transformation is to explicitly write it as part of the code-base.
However, \qs~is able to employ powerful, and type-safe, symbolic computation on the body of the given operation to automatically generate the adjoint if it is logically possible to do so.
This is, in effect, an algebra over the set of operations, and different variants can be automatically generated and combined for each given operation.
A host program calling into a \qs~component can be executed against a simulator (which provides an implementation for each of the primitive quantum gates), or translated to primitive physical operations which then run against appropriate hardware.

\qs is part of Microsoft's Quantum Development Kit, available at \url{http://www.microsoft.com/quantum}. Detailed documentation for \qs, including the standard library reference, is available at \url{http://docs.microsoft.com/quantum}.

%=============================================================================
\section{Related Work}
\label{sec:related}
%=============================================================================

A handful of quantum programming languages have been proposed in recent years, ranging from imperative to functional and low-level to high-level~\cite{Miszczak2011}.
Languages such as Quipper \cite{Selinger2013}, ScaffCC/Scaffold \cite{Scaffold,ScaffoldASPLOS}, \LIQUID  \cite{liquid}, QWire \cite{PRZ:2017}, Quil \cite{smith_practical_2016}, and ProjectQ \cite{PQ} enable programming of quantum computations and are intended as {\it circuit definition} languages.
Quipper is a strongly-typed, functional quantum programming language embedded in Haskell; ScaffCC/Scaffold is embedded in C/C++, leveraging the LLVM infrastructure; QWire is embedded in the proof system Coq; \LIQUID is embedded in F\#; and ProjectQ and Quil are embedded in Python.
Each of these examples offers powerful and extensible facilities for quantum circuit description and manipulation, including gate decomposition and circuit optimization; classical components such as measurements and classically-controlled gates; circuits at multiple levels of abstraction; exporting of quantum circuits for rendering or resource costing; and all systems are modular and user-extensible.

The abstractions and core features, however, focus on creating and manipulating quantum circuits.
While several languages, such as Quipper, \LIQUID, and ProjectQ, support higher-level functions, they are implemented as circuit transformations: functions that take an input circuit(s) and produce an output circuit(s).
This prevents these DSLs from modeling, for example, repeat-until-success algorithms and other algorithms with non-trivial branching.
This limitation can be partially mitigated by including some kinds of classical feedback as circuit elements, as is done at a high-level by \LIQUID or ProjectQ and at a low-level by OpenQASM 2.0 \cite{cross_open_2017}.
Including classical feedback directly in circuit representations is impractical for interacting robust classical algorithms with quantum processing, limiting the utility for adaptive characterization or for hybrid quantum--classical computations.

In contrast, \qs~is an {\it algorithm definition} language, and as such naturally represents the composition of classical and quantum algorithms alike.
There is no notion of a circuit in \qs, and there are \qs~statements, e.g., repeat-until-success, that cannot be represented as a circuit without introducing new and specialized gates and are rather most easily treated as hybrid quantum--classical constructs.
For instance, \qs~makes it easy to express phase estimation \cite{svore_faster_2013,wiebe_efficient_2016,kimmel_robust_2015} and quantum chemistry \cite{reiher_elucidating_2017} algorithms, both of which require rich quantum--classical interactions.

Another difference between \qs~and the other mentioned DSLs embedded in classical languages such as C++ \cite{mccaskey_extreme-scale_2017} or Python \cite{PQ}, \qs~uses a type model designed from the ground up, emphasizing classical determinism, first-class callables, and opacity of qubit types.
In contrast to the embedded quantum DSLs, \qs~is a bone fide stand-alone language. This means that functions provided by a large host language are not available; however, it has the key advantage that limitations of the host language do not apply to \qs~itself.

%=============================================================================
\section{Quantum Computing}
%=============================================================================

A quantum computer is a machine that stores and processes quantum information~\footnote{For an introduction to quantum computing, please refer to  \cite{Nielsen2000}.}.
The basic unit of quantum information is the quantum bit, also called \emph{qubit} for short.
The state of a qubit is a normalized vector in a two-dimensional Hilbert space: $\ket{\psi} = \alpha\ket{0} + \beta\ket{1}$, where $\alpha,\beta\in \mathbb{C}$ and $|\alpha|^2 + |\beta|^2=1$.
A note on terminology: the components $\alpha$, $\beta$ are called \emph{amplitudes} and a state of the above form is called a \emph{superposition} of the basis states $\ket{0}$ and $\ket{1}$, provided that $\alpha \ne 0 \ne \beta$.
The states corresponding to $\{0,1\}$ are often called {\it computational basis states}.
Their vector representation is given by $\ket{0} = \begin{bmatrix}\setlength{\arraycolsep}{3pt} 1 & 0 \end{bmatrix}^T$, and $\ket{1} = \begin{bmatrix} 0 & 1 \end{bmatrix}^T$, respectively.
Alternatively, a qubit can be viewed as a point on the surface of the so-called {\it Bloch sphere}, where $\ket{\psi} = \cos\frac{\theta}{2}\ket{0} + e^{i\phi}\sin\frac{\theta}{2}\ket{1}$, with $0 \leq \theta \leq \pi$ and $0\leq\phi<2\pi$.
This geometric representation represents a state by means of spherical coordinates with $\theta$ as the colatitude with respect to the $Z$-axis and $\phi$ as the longitude with respect to the $Y$ axis.
One advantage of the Bloch sphere representation is that each state is represented by a point on the unit sphere in $\mathbb{R}^3$ with $X, Y, Z$ axes.

To extract classical information, {\it measurements} can be performed.
To measure for instance the quantum state $\ket{\psi}$ in the computational basis, either state $\ket{0}$ or $\ket{1}$ is observed with probability $|\alpha|^2$ or $|\beta|^2$, respectively.
Measurements can be made along an arbitrary axis of the Bloch sphere, not just in the computational basis (the $Z$-axis).

More generally, the state vector of $n$ qubits lives in a $2^n$-dimensional Hilbert space and is represented by a complex vector $\ket{\psi} = \sum_{i=0}^{2^n-1} \alpha_i \ket{i}$ where $\sum_i |\alpha_i|^2=1$ and $i$ is an $n$-bit integer.
As an example, the four-qubit state $\ket{0000}$ is equivalent to the 4-way tensor product: 
$\ket{0}\otimes\ket{0}\otimes\ket{0}\otimes\ket{0} = \ket{0}^{\otimes 4} = \left[\begin{smallmatrix} 1 & 0 & 0 & \ldots & 0 \end{smallmatrix}\right]^T$, which as a state vector consists of a $1$, followed by 15 consecutive $0$s.
    This is often also written as $\ket{0,0,0,0}$ or $\ket{0}\ket{0}\ket{0}\ket{0}$.

Superpositions can extend over exponentially many states, while the underlying hardware scales only with a linear number of qubits.
This is an essential ingredient of a quantum algorithm, providing an innate massive parallelism within the quantum computer.

Finally, we note that computation evolves \emph{unitarily}, such that quantum operations must be \emph{reversible}, with state measurement as the exception as it collapses the quantum state to an observed value,  erasing knowledge of amplitudes $\alpha_i$.
If a quantum algorithm acts on $n$ qubits and if it does not involve intermediate measurements it can be modeled by a $2^n \times 2^n$ unitary matrix $U$ that acts on an $n$-qubit quantum state. Being unitary means that $UU^\dagger=I$, where $\dagger$ represents the adjoint operation (the complex conjugate transpose).
We introduce some special gates next that are important building blocks in quantum computing and therefore are part of the primitive \qs~library. First, there is the so-called {\it Hadamard} operation \lstinline+H+ which maps:
$\ket{0} \rightarrow \frac{1}{\sqrt{2}}\left( \ket{0} + \ket{1}\right)$, and
$\ket{1} \rightarrow \frac{1}{\sqrt{2}}\left( \ket{0} - \ket{1}\right)$.
Next, there are the Pauli \lstinline+X+ operation, which acts like a classical NOT gate in that it maps:
$\ket{0}\rightarrow \ket{1}$, and
$\ket{1}\rightarrow \ket{0}$.
Next, there are the Pauli \lstinline+Z+ gate which maps $\ket{1}\rightarrow -\ket{1}$ and the identity gate which is represented by \lstinline+I+. Finally, there is the \lstinline+T+ operation rotates a qubit around the $Z$-axis by angle $\pi/8$.
We also need a two-qubit operation for which the so-called {\it controlled-NOT} operation \lstinline+CNOT+ is a convenient choice. It maps $\ket{x,y}\rightarrow\ket{x, x\oplus y}$.
The corresponding unitary matrices are as follows:
\vspace{-1mm}
 \[
\lstinline+H+ =
  \frac{1}{\sqrt{2}}\left[\begin{smallmatrix}1&1\\1&\textrm{-}1\end{smallmatrix}\right],
  ~ \lstinline+X+ =
  \left[\begin{smallmatrix}0&1\\1&0\end{smallmatrix}\right], 
~ \lstinline+Z+  =
  \left[\begin{smallmatrix}1&0\\0&\textrm{-}1\end{smallmatrix}\right],
~ \lstinline+I+ = \left[\begin{smallmatrix}1&0\\0&1\end{smallmatrix}\right],
  \lstinline+T+ = \left[\begin{smallmatrix}1&0\\0&e^{i\pi/4}\end{smallmatrix}\right],
~ \mbox{and \lstinline+CNOT+} = \left[\begin{smallmatrix}1 & 0 & 0 & 0 \\ 0 & 1 & 0 & 0 \\ 0 & 0 & 0 & 1 \\ 0 & 0 & 1 & 0\end{smallmatrix}\right].
\]
The unitary matrices for multiple qubit versions of the above can be computed by appropriate tensor products with the identity operator $\lstinline+I+ = \left[\begin{smallmatrix}1 & 0 \\ 0 & 1\end{smallmatrix}\right]$.
There are also operations on more than $2$ qubits that are important. Among these is the so-called Toffoli gate, a controlled-controlled-NOT operator denoted \lstinline+CCNOT+ that maps $\ket{x, y, z}\rightarrow \ket{x, y, xy \oplus z}$.
Together \lstinline+H+, \lstinline+T+, and \lstinline+CNOT+, commonly called the Clifford + $T$ set, represent a {\it universal} set in that they can be used to express any quantum algorithm, along with measurement, to arbitrary accuracy (see, e.g., Refs.~\cite{KMM:2016,Selinger,BRS:2015} and references therein).

%-----------------------------------------------------------------------------
\subsection{Quantum Model of Computation}
%-----------------------------------------------------------------------------

A natural model for quantum computation is to treat the quantum computer as a coprocessor, similar to GPUs, FPGAs, and other adjunct processors.
The primary control logic runs classical code on a classical ``host'' computer.
When appropriate and necessary, the host program can invoke a sub-program that runs on the adjunct quantum processor.
When the sub-program completes, the host program gets access to the sub-program's results.

In this model, there are three levels of computation:
\begin{itemize}
    \item Classical computation that reads input data, sets up the quantum computation, triggers the quantum computation, processes the results of the computation, and presents the results to the user.
    \item Quantum computation that happens directly in the quantum device and implements a quantum algorithm.
    \item Classical computation that is required by the quantum algorithm during its execution.
\end{itemize}

There is no intrinsic requirement that these three levels all be written in the same language. 
Indeed, quantum computation has somewhat different control structures and resource management needs than classical computation, 
so using a custom programming language allows common patterns in quantum algorithms to be expressed more naturally.
At the same time, many quantum algorithms require intermediate classical computations during the execution of the algorithm,
so we have ensured that \qs~has the ability to express such classical computations as well. 

Alternatively, an embedded language brings with it all of the functionality of the host language. 
While this has some clear advantages it also has disadvantages when attempting to automatically perform meta operations, 
such as reversing a quantum circuit or conditioning the execution. 
By developing \qs~from the ground up, we ensure it is designed explicitly for easy expression, 
compilation, and optimization of quantum algorithms working in concert with classical computation.

%-----------------------------------------------------------------------------
\subsection{Quantum Algorithm Design}
%-----------------------------------------------------------------------------

Quantum algorithm design is both challenging and rewarding, with
many quantum algorithms achieving speedups over their classical counterparts \cite{JordanZoo}. 
At the start of a quantum algorithm, the qubit states are each initialized to $\ket{0}$.
A series of quantum operations are applied to evolve the quantum state. Additional qubits, called {\it ancilla}, may be used as "scratch space''.  Clean ancilla are initialized to $\ket{0}$ states and must be reset to $\ket{0}$ to allow later reuse.  
\qs~also allows to borrow qubits in an unknown state to serve as ancilla, provided they are returned to the same unknown state after use.

Common quantum subroutines to achieve speedups include {\it amplitude amplification} for increasing the amplitude of a desired state in a quantum superposition, {\it quantum phase estimation} for estimating eigenvalues of a unitary operator, and the {\it quantum Fourier transform} for performing a change of basis analogous to the classical discrete Fourier transform. 
The efficiency of the quantum Fourier transform (QFT) far surpasses what is possible on a classical machine making it one of the tools of choice when designing a quantum algorithm.
\qs~provides extensive built-in libraries containing implementations of these and other common subroutines, including the approximate quantum Fourier transform \cite{coppersmith_approximate_2002,RB:2008}, and a general version of oblivious amplitude amplification \cite{Berry2013}, making complex algorithm development simple and accessible.

Quantum algorithms may also require implementation of a {\it quantum oracle} for function evaluation. 
Reading and storing classical input data requires time linear in the amount of data and is often done through an oracle.  If a quantum algorithm achieves a quadratic computational speedup over its classical counterpart, the speedup may be lost once the oracle implementation is accounted for due to the required linear input time, thus the oracle implementation and optimization is especially important.
Classically, an oracle is a boolean function mapping an $n$-dimensional boolean input to an $m$-dimensional boolean output.
A classical boolean oracle can be converted into a quantum oracle by increasing the input and output spaces from $n$ and $m$ bits respectively to $n + m$ qubits each, enabling representation as a unitary matrix.
Example oracles include arithmetic functions, graph functions, and lookup tables. 
\qs~supports the definition of quantum oracles and provides a type-safe implementation of common oracles in its standard library.
Oracles are modeled in \qs~by operations that accept qubit arrays as a part of their input. 

To read the algorithm output, the final quantum state is measured.
Measurement presents an opportunity to extract information from the quantum computer. 
At the same time it presents a challenge since measurement is probabilistic and returns only $n$ bits of information upon measurement of the $2^n$ states in the $n$-qubit quantum state; measurement selects one output state at random among the $2^n$ states.
Evolving under a carefully constructed operation, such as the quantum Fourier transform or amplitude amplification, enables quantum interference to modify the states' amplitudes and in turn the probability distribution from which one draws the measurement output since Born's rule \cite{born_1926} provides that the absolute value of the amplitude squared maps to probability space.

To achieve dramatic quantum speedups, quantum algorithm designers should: (1) find problems requiring only a small amount of classical input; (2) seek an exponential speedup over the best-known classical algorithm; (3) leverage interference or amplitude amplification to sample from a target output distribution; and (4) use an efficient classical algorithm to post-process data. 

%=============================================================================
\section{A Taste of \qs}
%=============================================================================

In \autoref{lst:approximateqft}, we present a snippet of \qs~code from one of the published samples that are part of the \qs~software development kit.
This particular snippet defines a top-level operation that applies the Approximate Quantum Fourier Transform (AQFT) to a quantum register, see also 
\cite{coppersmith_approximate_2002,RB:2008} for more background on this operation. 

\begin{lstlisting}[
    caption={A \qs~implementation of the approximate quantum Fourier transform.},
    label={lst:approximateqft}
]
    namespace Microsoft.Quantum.Canon {
        open Microsoft.Quantum.Primitive;

        operation ApproximateQFT ( a: Int, qs: BigEndian) : () {
            body {
                let nQubits = Length(qs);

                for (i in 0 .. (nQubits - 1) ) {
                    for (j in 0..(i-1)) {
                    if ( (i-j) < a ) {
                        (Controlled R1Frac)( [qs[i]], (1, i - j, qs[j]) );
                        }
                    }
                    H(qs[i]);
                }

                // Apply the bit reversal permutation
                // to the quantum register
                SwapReverseRegister(qs);
            }

            adjoint auto
            controlled auto
            controlled adjoint auto
        }
    }
\end{lstlisting}

As you can see, \qs~is structurally very similar to familiar languages such as C\# and Java in its use of semicolons to end statements,
curly brackets to group statements, and double-slash to introduce comments.
\qs~also uses namespaces to group definitions together, and allows references to elements from other namespaces.

The basic unit in \qs~is an operation, which is essentially a function that can affect the state of the quantum device.
An operation can mix both classical and quantum computation.
Line 4 defines a new operation, \lstinline+ApproximateQFT+, that takes a pair of parameters and that has no return value.
When compiled, \lstinline+ApproximateQFT+ can be called from a classical host in a similar fashion to how a program might call a compiled GPU kernel.

The \lstinline+body+ element on line 5 specifies the implementation of the operation.
\qs~operations may also specify implementations for variants, or derived operations, that are common in quantum computing.
For instance lines 22, 23 and 24 indicate the parser this operation has an \lstinline+adjoint+ (inverse), a \lstinline+controlled+ and \lstinline+controlled adjoint+ variants that the parser should auto-calculate based on the body of the operation.

Operations can invoke other operations, like in lines 14 and 19 where the \lstinline+H+ and \lstinline+SwapReverseRegister+ operations are invoked; or their variants, like in line 11 where the \lstinline+controlled+ variant of the \lstinline+R1Frac+ operation is called.

\qs~offers classical flow and control constructs like in lines 8 and 9 in which the code iterates through a range of integers using \lstinline+for+, or \lstinline+if+\textendash\lstinline+elif+\textendash\lstinline+else+ to control execution like in line 10.
The control flow from the classical host through these control flow constructs and calls to other \qs~operations can be examined by the use of a debugger.

The \lstinline+let+ statements on line 6 create new immutable variable bindings.
\qs~has a functional flavor: variables are immutable by default, and the compiler performs type inference for all local variables.
Mutability is supported to help users more familiar with imparative languages, but is always isolated to the scope in which a variable is defined, as there are no explicit reference types in \qs.
Also, operations and functions can be used as return value or arguments to other operations, for example:

\begin{lstlisting}
    function OperationPow<`T>(oracle:(`T => ()), power : Int) : (`T => ())
    {
        return OperationPowImpl(oracle, power, _);
    }
\end{lstlisting}

This example also shows other unusual \qs~features: functions, type-parameters and partial application.

Functions are similar to operations, except they can only perform classical computation so they can't do or call any operations that will affect the state of a \lstinline+Qubit+ nor they can have any variants.
Thus, a classical host program calling into \qs~will typical begin by calling a \qs~operation.

The definition of an operation or a function may specify one or more type parameters, and then use those as the types of
the operation or function input or output parameters, like in line 1 in which \lstinline+OperationPow+ takes the type parameter \lstinline+`T+.

In \qs, you can create a new operation (or function) from an existing operation by providing a subset of the operation's parameters, like in line 3, in which the return value is a partial application of \lstinline+OperationPowImpl+.
This is similar to currying in other functional languages, but \qs~allows any subset of the parameters to be left unspecified, not just a final sequence.

%=============================================================================
\section{The \qs~Type System}
\label{sec:types}
%=============================================================================

\qs's type system starts with a familiar set of primitives, but provides some unusual ways to put them together to make new types.
In this section we'll walk through the primitive types and the mechanisms \qs~provides for creating more complex structures.

%-----------------------------------------------------------------------------
\subsection{Classical Primitives}
%-----------------------------------------------------------------------------

\qs~provides the standard classical data types \lstinline+Int+, \lstinline+Double+, \lstinline+Boolean+, and \lstinline+String+.
\lstinline+Int+, \lstinline+Double+, and \lstinline+Boolean+ all support the common operators on these types, such as arithmetic on numerics and logical operations on Booleans.
\qs~includes support for bit-wise operations on Ints since they are commonly used in quantum algorithms.

\lstinline+Strings+ in \qs~are used only for logging, and so minimal functionality is provided, reflecting that \qs~is intended to express computation running on an adjunt quantum processor.
Rather than providing a full string manipulation library, \qs~allows for debugging information to be passed back to a host program using C\#-style string interpolation; e.g. \lstinline+Message($"{measResult}")+.

\qs~also provides a \lstinline+Range+ type that represents an arithmetic sequence of integers.
For instance, the \lstinline+Range+ \lstinline+1..4+ is the sequence 1, 2, 3, 4, and the \lstinline+Range+ \lstinline+4..-1..1+ is the sequence 4, 3, 2, 1.
Ranges are first-class values that can be passed as parameters and as function or operation return values.

%-----------------------------------------------------------------------------
\subsection{Quantum Primitives}
%-----------------------------------------------------------------------------

\qs~provides three types that are useful in quantum computation: \lstinline+Pauli+, \lstinline+Result+ and \lstinline+Qubit+.

Values of the \lstinline+Pauli+ type specify a single-qubit Pauli operator; the possibilities are \lstinline+PauliI+, \lstinline+PauliX+, \lstinline+PauliY+, and \lstinline+PauliZ+.
Pauli values are used primarily to specify the basis for a measurement.

The \lstinline+Result+ type specifies the result of a quantum measurement.
\qs~mirrors most quantum hardware by providing measurements in products of single-qubit Pauli operators;
a \lstinline+Result+ of \lstinline+Zero+ indicates that the $+1$ eigenvalue was measured, and a \lstinline+Result+ of \lstinline+One+  indicates that the $-1$ eigenvalue was measured.
That is, \qs~represents eigenvalues by the power to which $-1$ is raised, using that $+1 = (-1)^0$ and $-1 = (-1)^1$.
This convention is more common in the quantum algorithms community, as it maps more closely to classical bits.

%~~~~~~~~~~~~~~~~~~~~~~~~~~~~~~~~~~~~~~~~~~~~~~~~~~~~~~~~~~~~~~~~~~~~~~~~~~~~~
\subsubsection{Qubits}
%~~~~~~~~~~~~~~~~~~~~~~~~~~~~~~~~~~~~~~~~~~~~~~~~~~~~~~~~~~~~~~~~~~~~~~~~~~~~~

\qs~treats qubits as opaque items that can be passed to both functions and operations, but that can only be interacted with by passing them to primitive (built-in) operations.
In particular, there is no type or construct in \qs~that represents the quantum state.
Instead, a qubit represents the smallest addressable physical unit in a quantum computer.
As such, a qubit is a long-lived item, so \qs~has no need for linear types. 

Importantly, we do not explicitly refer to the state within \qs, 
but rather describe how the state is transformed by the program. 
Similar to how a graphics shader program accumulates a description of transformations to each vertex, 
a quantum program in \qs~accumulates transformations to quantum states, 
represented as entirely opaque reference to the internal structure of a target machine. 

A \qs~program has no ability to introspect into the state of a qubit, 
and thus is entirely agnostic about what a quantum state is or on how it is realized. 
Rather, a program can call operations such as \lstinline+Measure+ to learn information from a qubit,
and operations such as \lstinline+X+ and \lstinline+H+ to act on the state of a qubit. 
Although \qs~defines a standard set of such primitve operations, these operations have no intrinsic definition within the language, 
and are made concrete only by the target machine used to run a particular \qs~program.
A \qs~program combines these operations as defined by a target machine to create new, 
higher-level operations to express quantum computation. 
In this way, \qs~makes it very easy to express the logic underlying quantum and hybrid quantum--classical 
algorithms, while also being very general with respect to the structure of a target machine and its
realization of quantum state.
That functions cannot modify the state of qubits passed as input arguments is then enforced by the restriction that functions can only call other functions, and cannot call operations.

%~~~~~~~~~~~~~~~~~~~~~~~~~~~~~~~~~~~~~~~~~~~~~~~~~~~~~~~~~~~~~~~~~~~~~~~~~~~~~
\subsubsection{Ancilla management and dirty qubits}
%~~~~~~~~~~~~~~~~~~~~~~~~~~~~~~~~~~~~~~~~~~~~~~~~~~~~~~~~~~~~~~~~~~~~~~~~~~~~~

In \qs, an algorithm always starts with no qubits and allocates and releases as it goes.
In this regard, \qs~models the quantum computer as a qubit heap.
Rather than supporting separate allocate and free statements or functions, \qs~has a \lstinline+using+ statement that allocates an array of qubits,
executes a block of statements, and releases the qubits at the end.
The using statement ensures that qubits don't get allocated and never freed or get released twice, thus avoiding a large class
of bugs common in manual memory management languages without the overhead of qubit garbage collection.

Some quantum algorithms are capable of using ``dirty ancillas;'' that is, they require extra qubits temporarily, but they can ensure that those qubits
are returned exactly to their original state by the end of the operation.
This means that, if there are qubits that are in use but not touched during the execution of a sub-algorithm, those qubits can be borrowed
for use as dirty ancillas instead of having to allocate new, clean ancillas.
Borrowing instead of allocating can significantly reduce the overall qubit requirements of an algorithm, and is expressed in \qs~by the \lstinline+borrowing+ statement.

%-----------------------------------------------------------------------------
\subsection{Collections}
%-----------------------------------------------------------------------------

\qs~offers two ways to collect multiple values into one entity: arrays and tuples.
Both tuples and arrays support elements of any \qs~type, including primitives, arrays, tuples, and operations.

An array is an ordered sequence of elements that all have the same type, similar to arrays in other languages.
As usual, array elements are accessed by index, and fetching an element by index is a constant-time operation.
\qs~does not support multi-dimensional arrays, but it does support arrays of arrays (i.e., jagged arrays).

\qs~arrays have two features that are more common for list types: concatenation and slicing.
Array concatenation forms a new array from two existing arrays by effectively gluing them together.
Array slicing forms a new array from an existing array and a range by using the elements of the range as indices into the array.
Among other uses, array slicing allows easy order inversion of an array by slicing with the range that counts from the last index to zero by negative one.

A tuple is an ordered collection of elements that can have different types.
In \qs, tuples once created may not be modified.
Tuples may have any number of elements, including zero and one, corresponding to the empty and singleton tuples, respectively.

Tuple elements may not be accessed by index; rather, tuple elements are accessed using a deconstructing let statement.
For instance, the statement:
\begin{lstlisting}
    let (a, b, (c, d), e) = (1, One, (1..3, PauliX), ("hello", [1; 5; 3]));
\end{lstlisting}
results in binding \lstinline+a+ to \lstinline+1+, \lstinline+b+ to \lstinline+One+, \lstinline+c+ to \lstinline+1..3+, \lstinline+d+ to \lstinline+PauliX+, and \lstinline+e+ to \lstinline+("hello", [1; 5; 3])+.

%%~~~~~~~~~~~~~~~~~~~~~~~~~~~~~~~~~~~~~~~~~~~~~~~~~~~~~~~~~~~~~~~~~~~~~~~~~~~~~
%\subsubsection{Singleton Tuple Equivalence}
%%~~~~~~~~~~~~~~~~~~~~~~~~~~~~~~~~~~~~~~~~~~~~~~~~~~~~~~~~~~~~~~~~~~~~~~~~~~~~~

In \qs, a tuple value with one element is considered identical to the ``unwrapped'' element, with the same holding for types.
For instance, the types \lstinline+Int+ and \lstinline+(Int)+ are treated as identical by \qs, as are the values \lstinline+5+, \lstinline+(5)+ and \lstinline+(((5)))+.
This is known as \emph{singleton tuple equivalence}, and allows us to simplify many areas of \qs.
Since all types in \qs~are resolvable at compile-time (there is no dynamic dispatch or reflection), singleton tuple equivalence can be readily implemented during compilation.
In particular, singleton tuple equivalence allows us to specify that all operations and functions in \qs~both accept and return tuples.

%-----------------------------------------------------------------------------
\subsection{Operations and Functions}
%-----------------------------------------------------------------------------

\qs~supports two types of callables: operations and functions.
Operations, as we've seen above, are routines that can affect the quantum state.
Functions are routines that are guaranteed not to affect the quantum state and that will always return the
same result, given the same inputs.
For instance, a function may perform a classical computation that is required by the quantum algorithm.

All operations and functions take a tuple as input and produce a tuple as output.
Those that produce no result return the empty tuple, \lstinline+()+. Because of singleton tuple equivalence, a function or operation that returns a singleton tuple may be treated as returning a single
unwrapped value.

Operations and functions are first-class types in \qs.
They can be passed in as parameters to other operations and functions, and may be returned as the result of an operation or function.
Functions and operations may be included in tuples, and you can have arrays of operations or functions.

%~~~~~~~~~~~~~~~~~~~~~~~~~~~~~~~~~~~~~~~~~~~~~~~~~~~~~~~~~~~~~~~~~~~~~~~~~~~~~
\subsubsection{Variants and Functors}
%~~~~~~~~~~~~~~~~~~~~~~~~~~~~~~~~~~~~~~~~~~~~~~~~~~~~~~~~~~~~~~~~~~~~~~~~~~~~~

In quantum computing, most operations are unitary, which is to say they have an inverse that is mathematically
equal to their complex conjugate transpose, or adjoint.
The adjoint of an operator exactly undoes the quantum state change performed by the original operation.
Many quantum algorithms require both an operation and its adjoint in order to perform a computation.
Similarly, many operations may be controlled on the state of a quantum register.
The controlled variant of an operation takes one or more additional qubits, and either performs the operation on the remaining inputs
if all of the control qubits are in the \lstinline+One+ state, or does nothing if the control qubits are in any other combination of states.
The controlled variant is more than just wrapping the operation in an if statement, since it does this coherently: 
if the control qubits are in a superposition of states, then the result is a superposition of both applying and not applying the operation.

\qs~allows an operation definition to specify whether it has either an adjoint or a controlled variant, or both.
Operations that have an adjoint may specify that they are their own inverse (that is, they are ``self-adjoint''), may provide
the implementation of their adjoint, or may request the compiler to generate an adjoint for them.
This last option is not available for all operations, but many operations that appear in common algorithms allow automatic adjoint generation.
Similarly, operations that have a controlled variant may either provide the implementation of their controlled variant
or may request the compiler to generate the variant.

Operations that have both a controlled variant and an adjoint must also specify the result of applying both variations; that is,
they must define their controlled adjoint variant by either providing an implementation or requesting one to be generated.

The following is an example of of the \lstinline+CCNOT+ operation in \qs~which is self-adjoint and has a controlled variant that will be auto-generated by the compiler based on the body definition:
\begin{lstlisting}
    operation CCNOT (control1 : Qubit, control2 : Qubit, target : Qubit)  : ()
    {
        body 
        {
            (Controlled X)([control1; control2],target);
        }
        adjoint self
        controlled auto
        adjoint controlled auto    
    }
\end{lstlisting}

In use, if \lstinline+op+ is an operation that has an adjoint, \lstinline+(Adjoint op)+ is the operation that is the adjoint of \lstinline+op+,
and \lstinline+(Controlled op)+ is the controlled version of \lstinline+op+.
In these constructs, \lstinline+Adjoint+ and \lstinline+Controlled+ are functors: factories that define a new operation from another operation.
Functors in \qs~have access to the implementation of the base operation when 
defining the implementation of the new operation. 
Thus, functors can perform more complex functions than 
traditional higher-level functions.

%~~~~~~~~~~~~~~~~~~~~~~~~~~~~~~~~~~~~~~~~~~~~~~~~~~~~~~~~~~~~~~~~~~~~~~~~~~~~~
\subsubsection{Partial Application}
%~~~~~~~~~~~~~~~~~~~~~~~~~~~~~~~~~~~~~~~~~~~~~~~~~~~~~~~~~~~~~~~~~~~~~~~~~~~~~

As mentioned above, new operations and functions may be created at run-time by supplying some but not all of the parameters of the operation or function.
The missing parameters are designated by an underscore, \lstinline+_+.
The resulting value is an operation or function that takes only the missing parameters.
If the base operation has an adjoint or controlled variant, then the result does as well.

Because partial application of an operation does not actually evaluate the operation, it has
no impact on the quantum state.
This means that building a new operation from existing operations and computed data may be done in a function;
this is useful in many adaptive quantum algorithms and in defining new flow control constructs.

\qs~allows very general partial application for operations and functions that take complex tuples as input.
For instance, given an operation \lstinline+Op+ whose input is the tuple type \lstinline+(Int, (Double, Qubit), Int)+,
\lstinline+Op(1, (1.0, _), _)+ is a partial application whose result is an operation that takes a \lstinline+((Qubit), Int)+ input.
By singleton tuple equivalence, this is the same as taking a \lstinline+(Qubit, Int)+ input.

%~~~~~~~~~~~~~~~~~~~~~~~~~~~~~~~~~~~~~~~~~~~~~~~~~~~~~~~~~~~~~~~~~~~~~~~~~~~~~
\subsubsection{Generics}
%~~~~~~~~~~~~~~~~~~~~~~~~~~~~~~~~~~~~~~~~~~~~~~~~~~~~~~~~~~~~~~~~~~~~~~~~~~~~~

\qs~supports type-parameterized operations and functions, which are commonly referred to as generics.
The definition of an operation or function may specify one or more type parameters, and then use those as the types of
the operation or function's input or output parameters.
A type parameter may appear multiple times in the input and output types of an operation or function.

For example, the \lstinline+Map+ function from the standard library has two type parameters, \lstinline+`T+ and \lstinline+`U+.
It takes as arguments a function from \lstinline+`T+ to \lstinline+`U+ and an array of elements of type \lstinline+`T+,
and returns the array of elements of type \lstinline+`U+ generated by applying the function to each of the elements
in the input array in turn.

The type parameters of a generic operation or function are inferred when the operation or function is called.
It is possible to call the same generic operation or function with completely different arguments, even within a single expression.
It is also possible to pass a generic operation or function as a parameter to another operation or function.

When combined with partial application, generic operations and functions work as one would expect: all of the type parameters
that can be inferred from the provided parameters are set in the resulting operation, and any left free remain free.

The \qs~standard library makes heavy use of generics to provide a host of useful abstractions,
including functions like \lstinline+Map+ and \lstinline+Fold+ that are familiar from functional languages.

%-----------------------------------------------------------------------------
\subsection{User-Defined Types}
%-----------------------------------------------------------------------------

\# allows a new named type to be defined as equal to a type expression.
For instance, a user-defined type or UDT may be defined based on a tuple type or an operation signature.

In some respects, UDTs act as aliases for their base types, but have a bit more meaning than a simple alias.
A UDT is treated as a strict subtype of its base type, so that a value of a UDT type can be used where
a value of the base type is expected, but not vice versa.
If two UDTs share the same base type, a value of one UDT type may not be used where a value of the other UDT type is expected.

The standard \qs~library uses UDTs to specialize various base types.
For instance, the library includes quantum arithmetic operations for both big-endian and little-endian integers.
It defines two UDTs, \lstinline+BigEndian+ and \lstinline+LittleEndian+, both of which are based on \lstinline+Qubit[]+.
This allows operations to specify whether they are written for big-endian or little-endian representations,
and leverages the type system to ensure at compile-time that mismatched operands aren't allowed.

%=============================================================================
\section{\qs~Statements}
%=============================================================================

Most \qs~statements are familiar from languages such as C\# and Java.
Some statements, such as the \lstinline+let+ and \lstinline+mutable+ binding statements, are more similar to F\# and other functional languages, while others such as \lstinline+repeat+--\lstinline+until+ and \lstinline+borrowing+ are unique to \qs.

%-----------------------------------------------------------------------------
\subsection{Binding and Assignment}
%-----------------------------------------------------------------------------

\qs~has three statements that deal with variables:
\begin{itemize}
    \item \lstinline+let+ is used to create an immutable binding.
    \item \lstinline+mutable+ is used to create a mutable binding.
    \item \lstinline+set+ is used to change the value of a mutable binding.
\end{itemize}

In all cases, variables types are inferred from the type of the right-hand side of the binding.
It is illegal to change the type of a variable in a \lstinline+set+ statement.

For array variables, both the array itself and the item contents are either immutable or mutable, depending
on whether the variable was bound with \lstinline+let+ or \lstinline+mutable+.
That is, to create an array whose elements will be updated later, the array must be mutably bound.

Function and operation arguments are always immutably bound; there are no ``out'' arguments in \qs.
In particular, since the states of \lstinline+Qubit+ values are not defined or observable from within \qs, this does not preclude the accumulation of quantum side effects through calls to primitive operations.

%-----------------------------------------------------------------------------
\subsection{Flow Control}
%-----------------------------------------------------------------------------

\qs~has a very standard \lstinline+if+\textendash\lstinline+elif+\textendash\lstinline+else+ statement.
The \lstinline+if+ clause, and any \lstinline+elif+ clauses, take a Boolean test value; the body of the first clause whose test value is true is executed.
If no test value is true, then the body of the \lstinline+else+ clause is executed, if there is one.

The \lstinline+for+ statement in \qs~allows iteration through a \lstinline+Range+ value, with an integer loop variable.

A construct that is special for quantum computation is the \lstinline+repeat+\textendash\lstinline+until+\textendash\lstinline+fixup+ loop.
This loop executes the \lstinline+repeat+ body and then evaluates the \lstinline+until+ test value.
If the test is true, then the statement is complete.
Otherwise, the \lstinline+fixup+ body is executed, and the loop starts over.
This statement supports the ``repeat-until-success'' pattern in quantum computing.

The \lstinline+return+ statement exits from the current operation or function with a specified return value, and can be used anywhere other than in a scope which allocates qubits (that is, not from within a \lstinline+using+ or \lstinline+borrowing+ block).
An operation or function that has no result value by convention returns the empty tuple, \lstinline+()+.
%-----------------------------------------------------------------------------
\subsection{Qubit Management}
%-----------------------------------------------------------------------------

As mentioned above in the \lstinline+Qubit+ section, \qs~provides the \lstinline+using+ and \lstinline+borrowing+ statements for qubit management.
Both of these statements gather an array of qubits and bind it to a variable specified in the statement, and then execute a block of statements.
At the end of the block, the qubits are given back.

For the \lstinline+using+ statement, the qubits are allocated from the quantum computer's free qubit heap, and then returned to the heap.
For the \lstinline+borrowing+ statement, the qubits are allocated from in-use qubits that are guaranteed not to be used during the body of the statement,
and left in their original state at the end.
If there aren't enough qubits available to borrow, then qubits will be allocated from and returned to the heap.

%=============================================================================
\section{Debugging and Testing \qs~Programs}
%=============================================================================

Debugging quantum programs is challenging due to the nature of quantum computation,  probabilistic measurements on quantum hardware, and the exponential size of the quantum state space.
In contrast to other quantum programming languages, \qs~provides robust functionality to enable detailed debugging of quantum programs.

Since \qs~functions are deterministic, a function whose output type is the empty tuple \lstinline+()+ cannot ever be observed from within a \qs~program.
That is, a target machine can choose not to execute any function which returns \lstinline+()}+with the guarantee that this omission will not modify the behavior of any following \qs~code. 
This consequence makes functions a useful tool for embedding debugging and testing logic.

In particular, functions of this form can be used to represent diagnostic side effects. 
Consider a simple example:
\begin{lstlisting}
    function AssertPositive(value : Double) : () {
        if (value <= 0) {
            fail "Expected a positive number.";
        }
    }
\end{lstlisting}
The keyword \lstinline+fail+ indicates the computation should halt, raising an exception in the target machine running the \qs~program. 
By definition, a failure of this kind cannot be observed from within \qs, as no further \qs~code is run after a \lstinline+fail+ statement is reached. 
Thus, if we proceed past a call to \lstinline+AssertPositive+, we can be assured by the anthropic principle that its input was positive, even though we did not directly observe this fact.

Similarly, the primitive function \lstinline+Message+ has type \lstinline+String -> ()+, and allows emitting diagnostic messages. 
That a target machine observes the contents of the input to \lstinline+Message+ has no consequences observable from within \qs. 
A target machine may thus elide calls to \lstinline+Message+ by the same logic.

Building on these ideas, \qs~offers two especially useful assertions, both modeled as functions onto \lstinline+()}:+\lstinline+Assert+ and \lstinline+AssertProb+. 
\lstinline+Assert+ asserts that measuring the given quantum register in the given Pauli basis will always produce the given result. \lstinline+AssertProb+ asserts that such a measurement will produce the given result with the given probability.
On target machines which work by simulation we are not bound by the no-cloning theorem, which states that an arbitrary quantum state cannot be copied, and can perform such measurements without disturbing the register passed to such assertions. 
A simulator can then, similar to the \lstinline+AssertPositive+ function above, abort computation if the hypothetical outcome would not be observed in practice:
\begin{lstlisting}
    using (register = Qubit[1]) {
        H(register[0]);
        Assert([PauliX], register, Zero);
        // PauliX measures in (|+>, |->) basis
        // We know state is |+> w/o accessing it.
    }
\end{lstlisting}
On actual hardware, where we are constrained by physics, we of course cannot perform such assertions without measurements and thus (potentially) disturbing the state. There the \lstinline+Assert+ and \lstinline+AssertProb+ functions simply return \lstinline+()+ with no other effect.

%=============================================================================
\section{Conclusions}
%=============================================================================

We present \qs, a scalable, high-level, domain-specific language for quantum programming.
Unlike previous quantum programming languages, \qs~is a stand-alone language offering a high level of abstraction, informative error reporting, a strongly-typed design that offers guarantees on type safety, and an extensive list of quantum libraries, including modular arithmetic, Shor's algorithm for factoring and elliptic curve dlogs, and Hamiltonian simulation.
It also exhibits innovative language features such as general partial application and the ability to treat unresolved generics as first class.

\qs~empowers developers with distinct capabilities in quantum computing, including quantum control-flow constructs such as \lstinline+repeat+--\lstinline+until+, extensive standard libraries with common quantum functions and subroutines, support of functors such as adding control to a circuit and computing the adjoint of a circuit, guarantees on type safety, rich compilation and error reporting, and state-of-the-art target machines for simulation and resource estimation.
Further, \qs's memory management enables both clean and dirty ancilla usage for resource optimization in quantum algorithms.

Future extensions include new \qs~target machines: a Toffoli simulator (e.g., to debug and test arithmetic circuits) and a Clifford operation simulator (e.g., for use in error correction), additional high-level algorithmic subroutines in the \qs~standard libraries, and enhanced debugging of quantum programs through quantum state visualization methods and more comprehensive code profiling.

%% END MATTER %%%%%%%%%%%%%%%%%%%%%%%%%%%%%%%%%%%%%%%%%%%%%%%%%%%%%%%%%%%%%%%%

%merlin.mbs apsrev4-1.bst 2010-07-25 4.21a (PWD, AO, DPC) hacked
%Control: key (0)
%Control: author (8) initials jnrlst
%Control: editor formatted (1) identically to author
%Control: production of article title (0) allowed
%Control: page (0) single
%Control: year (1) truncated
%Control: production of eprint (1) enabled
%

\end{document}